\documentstyle[prd,aps,preprint]{revtex}      % For preprints, drafts
\input psfig
\begin{document}

\draft % makes pacs numbers print
\tighten % To save trees, please use this for the copy to be distributed!
%  FOR THE PREPRINT/CLNS
\preprint{\vbox{\hbox{CLNS 95/1338 \hfill}
                \hbox{CLEO 95--8  \hfill}
                \hbox{\today       \hfill}}}

\title{Search for Exclusive Charmless Hadronic B Decays}

\author{
D.M.~Asner, M.~Athanas, D.W.~Bliss, W.S.~Brower, G.~Masek,
and H.P.~Paar}
\address{
University of California, San Diego, La Jolla, California 92093}
\author{
J.~Gronberg, C.M.~Korte, R.~Kutschke, S.~Menary, R.J.~Morrison,
S.~Nakanishi, H.N.~Nelson, T.K.~Nelson, C.~Qiao, J.D.~Richman,
D.~Roberts, A.~Ryd, H.~Tajima, and M.S.~Witherell}
\address{
University of California, Santa Barbara, California 93106}
\author{
R.~Balest, K.~Cho, W.T.~Ford, M.~Lohner, H.~Park, P.~Rankin,
and J.G.~Smith}
\address{
University of Colorado, Boulder, Colorado 80309-0390}
\author{
J.P.~Alexander, C.~Bebek, B.E.~Berger, K.~Berkelman, K.~Bloom,
T.E.~Browder,%
\thanks{Permanent address: University of Hawaii at Manoa}
D.G.~Cassel, H.A.~Cho, D.M.~Coffman, D.S.~Crowcroft, M.~Dickson,
P.S.~Drell, D.J.~Dumas, R.~Ehrlich, R.~Elia, P.~Gaidarev,
M.~Garcia-Sciveres, B.~Gittelman, S.W.~Gray, D.L.~Hartill,
B.K.~Heltsley, S.~Henderson, C.D.~Jones, S.L.~Jones,
J.~Kandaswamy, N.~Katayama, P.C.~Kim, D.L.~Kreinick, T.~Lee,
Y.~Liu, G.S.~Ludwig, J.~Masui, J.~Mevissen, N.B.~Mistry, C.R.~Ng,
E.~Nordberg, J.R.~Patterson, D.~Peterson, D.~Riley, and A.~Soffer}
\address{
Cornell University, Ithaca, New York 14853}
\author{
P.~Avery, A.~Freyberger, K.~Lingel, C.~Prescott, J.~Rodriguez,
S.~Yang, and J.~Yelton}
\address{
University of Florida, Gainesville, Florida 32611}
\author{
G.~Brandenburg, D.~Cinabro, T.~Liu, M.~Saulnier, R.~Wilson,
and H.~Yamamoto}
\address{
Harvard University, Cambridge, Massachusetts 02138}
\author{
T.~Bergfeld, B.I.~Eisenstein, J.~Ernst, G.E.~Gladding,
G.D.~Gollin, M.~Palmer, M.~Selen, and J.J.~Thaler}
\address{
University of Illinois, Champaign-Urbana, Illinois, 61801}
\author{
K.W.~Edwards, K.W.~McLean, and M.~Ogg}
\address{
Carleton University, Ottawa, Ontario K1S 5B6
and the Institute of Particle Physics, Canada}
\author{
A.~Bellerive, D.I.~Britton, E.R.F.~Hyatt, R.~Janicek,
D.B.~MacFarlane, P.M.~Patel, and B.~Spaan}
\address{
McGill University, Montr\'eal, Qu\'ebec H3A 2T8
and the Institute of Particle Physics, Canada}
\author{
A.J.~Sadoff}
\address{
Ithaca College, Ithaca, New York 14850}
\author{
R.~Ammar, P.~Baringer, A.~Bean, D.~Besson, D.~Coppage, N.~Copty,
R.~Davis, N.~Hancock, S.~Kotov, I.~Kravchenko, and N.~Kwak}
\address{
University of Kansas, Lawrence, Kansas 66045}
\author{
Y.~Kubota, M.~Lattery, M.~Momayezi, J.K.~Nelson, S.~Patton,
R.~Poling, V.~Savinov, S.~Schrenk, and R.~Wang}
\address{
University of Minnesota, Minneapolis, Minnesota 55455}
\author{
M.S.~Alam, I.J.~Kim, Z.~Ling, A.H.~Mahmood, J.J.~O'Neill,
H.~Severini, C.R.~Sun, and F.~Wappler}
\address{
State University of New York at Albany, Albany, New York 12222}
\author{
G.~Crawford, R.~Fulton, D.~Fujino, K.K.~Gan, K.~Honscheid,
H.~Kagan, R.~Kass, J.~Lee, M.~Sung, C.~White, A.~Wolf,
and M.M.~Zoeller}
\address{
Ohio State University, Columbus, Ohio, 43210}
\author{
X.~Fu, B.~Nemati, W.R.~Ross, P.~Skubic, and M.~Wood}
\address{
University of Oklahoma, Norman, Oklahoma 73019}
\author{
M.~Bishai, J.~Fast, E.~Gerndt, J.W.~Hinson, T.~Miao, D.H.~Miller,
M.~Modesitt, E.I.~Shibata, I.P.J.~Shipsey, and P.N.~Wang}
\address{
Purdue University, West Lafayette, Indiana 47907}
\author{
L.~Gibbons, S.D.~Johnson, Y.~Kwon, S.~Roberts, and E.H.~Thorndike}
\address{
University of Rochester, Rochester, New York 14627}
\author{
T.E.~Coan, J.~Dominick, V.~Fadeyev, I.~Korolkov, M.~Lambrecht,
S.~Sanghera, V.~Shelkov, T.~Skwarnicki, R.~Stroynowski,
I.~Volobouev, and G.~Wei}
\address{
Southern Methodist University, Dallas, Texas 75275}
\author{
M.~Artuso, M.~Gao, M.~Goldberg, D.~He, N.~Horwitz, S.~Kopp,
G.C.~Moneti, R.~Mountain, F.~Muheim, Y.~Mukhin, S.~Playfer,
S.~Stone, and X.~Xing}
\address{
Syracuse University, Syracuse, New York 13244}
\author{
J.~Bartelt, S.E.~Csorna, V.~Jain, and S.~Marka}
\address{
Vanderbilt University, Nashville, Tennessee 37235}
\author{
D.~Gibaut, K.~Kinoshita, and P.~Pomianowski}
\address{
Virginia Polytechnic Institute and State University,
Blacksburg, Virginia, 24061}
\author{
B.~Barish, M.~Chadha, S.~Chan, D.F.~Cowen, G.~Eigen, J.S.~Miller,
C.~O'Grady, J.~Urheim, A.J.~Weinstein, and F.~W\"{u}rthwein}
\address{
California Institute of Technology, Pasadena, California 91125}

\author{(CLEO Collaboration)}

\date{\today}
\maketitle

\begin{abstract}
We have searched for two-body charmless hadronic
decays of $B$ mesons.  Final states include
$\pi\pi$, $K \pi$, and $KK$ with both charged and
neutral kaons and pions; $\pi\rho$, $K \rho$,
and $K^*\pi$;
and $K\phi$, $ K^*\phi$, and $\phi\phi$.
The data used in this analysis consist of 2.6~million
$B\bar{B}$~pairs produced at the
$\Upsilon(4S)$ taken with the CLEO-II detector at the Cornell Electron
Storage Ring (CESR).  We measure the branching fraction of the
sum of $B^0 \rightarrow \pi^+\pi^-$ and $B^0 \rightarrow K^+\pi^-$
to be $(1.8^{+0.6+0.2}_{-0.5-0.3}\pm0.2) \times 10^{-5}$.
In addition, we place upper limits on individual
branching fractions in the range from $10^{-4}$ to $10^{-6}$.

\end{abstract}


\narrowtext
\section{Introduction}
\begin{figure}
\centerline{\hbox{\psfig{figure=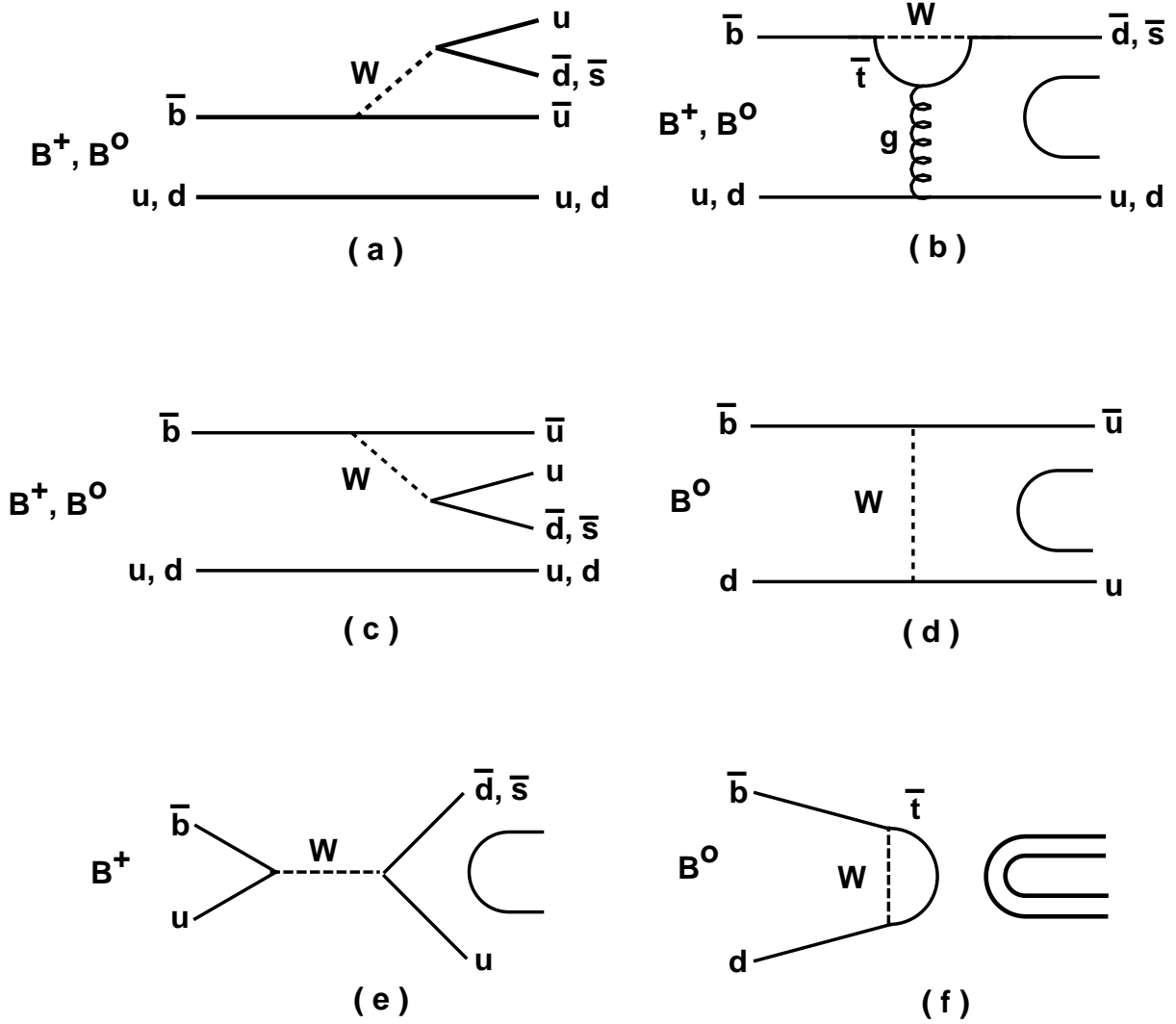,height=6in}}}
\vskip .5in
\caption{Feynman diagrams for rare hadronic $B$ decays:
(a) $b \rightarrow u$ external $W$ emission,
(b) $b \rightarrow s,d$ loop or gluonic penguin,
(c) $b\rightarrow u$ internal $W$ emission,
(d) $b\rightarrow u$ $W$ exchange,
(e) annihilation,
and (f) vertical $W$ loop.}
\label{fig:feynman}
\end{figure}

The decays of $B$ mesons to two charmless hadrons can be described by
a $b \rightarrow u$ tree-level spectator diagram
(Figure~\ref{fig:feynman}a),
or a $b \rightarrow sg$ one-loop ``penguin-diagram''
(Figure~\ref{fig:feynman}b) and to a lesser extent, by the
color-suppressed tree (Figure~\ref{fig:feynman}c)
or CKM-suppressed $b\rightarrow dg$ penguin diagrams.
Although such decays can also include contributions from
$b \rightarrow u$
$W$-exchange (Figure~\ref{fig:feynman}d), annihilation
(Figure~\ref{fig:feynman}e), or vertical $W$~loop
(Figure~\ref{fig:feynman}f) processes,
these contributions are expected to be
negligible in most cases.

Decays such as $B^0 \rightarrow \pi^+\pi^-$
and $B^0 \rightarrow \pi^\pm\rho^\mp$ are expected to be dominated by
the $b \rightarrow u$ spectator transition, and measurements of their
branching fractions could be used to extract a value for $|V_{ub}|$.
The decay mode $B^0 \rightarrow \pi^+\pi^-$ can be used to measure $CP$
violation in the $B$~sector at both asymmetric
$B$~factories~\cite{bfact} and hadron colliders\cite{hadcol}.
Since the $\pi^+\pi^-$ final state is a $CP$ eigenstate, $CP$
violation can arise from interference between the amplitude for
direct decay and the amplitude for the process in which the $B^0$
first mixes into a $\bar{B}^0$ and then decays.  Measurement
of the time evolution of the rate asymmetry leads to a measurement of
$\sin 2\alpha$, where $\alpha$ is one of the angles in the unitarity
triangle\cite{unitarity_triangle}.  If the
$B^0 \rightarrow \pi^+\pi^-$ decay has a non-negligible contribution
from the $b \rightarrow dg$ penguin,
interference between the spectator and penguin contributions
will contaminate the measurement of $CP$ violation via
mixing~\cite{pollution}, an effect
known as ``penguin pollution.''  If this is the case, the penguin and
spectator effects can be disentangled by also measuring the
isospin-related decays
$B^0\rightarrow \pi^0\pi^0$ and  $B^\pm\rightarrow \pi^\pm\pi^0$
\cite{pipicp}.
Alternatively,
SU(3) symmetry can be used to relate $B^0\rightarrow \pi^+\pi^-$ and
$B^0\rightarrow K^+\pi^-$ \cite{silva,oits566}.
Penguin and spectator effects may then
be disentangled \cite{silva}
once the ratio of the two branching fractions
and $\sin 2\beta$\cite{unitarity_triangle} are
measured.

Decays such as $B^0 \rightarrow K^+\pi^-$ and $B^0 \rightarrow K^{*+}\pi^-$
are expected to be dominated by the $b \rightarrow sg$ penguin process,
with a small contribution from a Cabibbo-suppressed $b \rightarrow u$ spectator
process.  Interference between the penguin and spectator amplitudes can
give rise to direct $CP$ violation, which will manifest itself
as a rate asymmetry for
decays of $B^0$ and $\bar{B}^0$~mesons, but the presence of
hadronic phases
complicates the extraction of the $CP$ violation parameters.

There has been discussion in recent literature about extracting the
unitarity angles using precise time-integrated measurements
of $B$ decay rates. Gronau,
Rosner, and London have proposed~\cite{kpicp} using isospin relations
and flavor SU(3) symmetry to extract, for example, the unitarity angle
$\gamma$ by measuring the rates of $B^+$ decays to
$K^0\pi^+$, $K^+\pi^0$, and $\pi^+\pi^0$ and their
charge conjugates.  More recent publications
\cite{deshpande,GRLSU3,EWPrebuttal,EWPmore} have questioned whether
electroweak penguin contributions
($b \rightarrow s\gamma$, $b\rightarrow sZ$)
are large enough to invalidate isospin relationships
and whether SU(3) symmetry-breaking effects can be
taken into account.
If it is possible to extract unitarity angles from rate measurements alone,
the measurements could be made at either symmetric or
asymmetric $B$~factories (CESR, KEK, SLAC),
but will
require excellent particle identification to distinguish between the
$K\pi$ and $\pi\pi$ modes.

Decays such as $B \rightarrow K\phi$ and $B^+ \rightarrow K^0\pi^+$
cannot occur via a spectator process and are expected to be
dominated by the penguin process.  Measurement of these decays will
give direct information on the strength of the penguin amplitude.

Various extensions or alternatives to the Standard Model have been suggested.
Such models characteristically involve hypothetical high mass particles,
such as fourth generation quarks, leptoquarks, squarks,
gluinos, charged Higgs, charginos, right-handed $W$'s, and so on.
They have negligible  effect on tree diagram dominated $B$ decays, such as
those involving $b\rightarrow cW^-$ and $b\rightarrow uW^-$, but can
contribute significantly to loop processes like $b\rightarrow sg$ and
$b\rightarrow dg$.

Since non-standard models can have
enhanced $CP$ violating effects relative to predictions based on
the standard Kobayashi-Maskawa mechanism \cite{RKM,RNirQuinn},
such effects
might turn out to
be the key to the solution of the baryogenesis problem, that is,
the obvious asymmetry in the abundance of baryons over antibaryons
in the universe.
Many theorists believe
that the KM mechanism
for $CP$ violation is not sufficient to generate the observed
asymmetry or even to maintain an initial asymmetry through
cool-down \cite{RcosmicCP}.
Loop processes
in $B$ decay may be our most sensitive probe of physics
beyond the Standard Model.

This paper reports results on the decays $B \rightarrow \pi\pi$,
$B \rightarrow K \pi$, $B\rightarrow KK$, $B\rightarrow
\pi\rho$, $B\rightarrow K \rho$,
$B\rightarrow K^*\pi$, $B \rightarrow K\phi$, $B \rightarrow K^*\phi$,
and $B \rightarrow \phi \phi$ \cite{chargeconjugation}.
Recent observations of the sum of the two-body charmless hadronic
decays $B^0\rightarrow \pi^+\pi^- {\rm and}\ K^+\pi^-$~\cite{kpi} and
of the electromagnetic penguin decay $B \rightarrow K^*
\gamma$~\cite{kstargamma}, indicate that we have reached the
sensitivity required to observe such decays.
The size of the data
set and efficiency of the
CLEO detector allow us to
place upper limits on the branching fractions in
the range $10^{-4}$ to $10^{-6}$.

\section{Data sample and event selection}
The data set used in this analysis was collected with the CLEO-II
detector~\cite{detector} at the Cornell Electron Storage Ring (CESR).
It consists of $2.42~{\rm fb}^{-1}$ taken at the $\Upsilon$(4S)
(on-resonance) and $1.17~{\rm fb}^{-1}$ taken at a center of mass energy
about 35~MeV below $B\bar{B}$ threshold.  The
on-resonance sample contains 2.6~million $B\bar{B}$ pairs.
The below-threshold sample
is used for continuum background estimates.

The momenta of charged particles are measured in a
tracking system consisting of a 6-layer straw
tube chamber, a 10-layer precision
drift chamber, and a 51-layer main drift chamber, all operating
inside a 1.5 T superconducting solenoid.  The main drift chamber
also provides a measurement of the specific ionization loss, $dE/dx$,
used for particle identification.
Photons are detected
using 7800 CsI crystals, which are also inside the magnet. Muons are
identified using proportional counters placed at various depths in the
steel return yoke of the magnet.
The excellent efficiency and resolution of the CLEO-II detector for
both charged particles and photons are crucial in
extracting signals and suppressing both continuum
and combinatoric backgrounds.

Charged tracks are
required to pass track quality cuts based on the average hit residual
and the impact parameters in both the $r-\phi$ and $r-z$ planes.
We require that charged track momenta be greater than 175~MeV/$c$ to
reduce low momentum combinatoric background.

Pairs of tracks with vertices displaced from the primary interaction
point are taken as $K_S^0$ candidates.  The secondary vertex is
required to be displaced from the primary interaction point by at
least 1~mm for candidates with momenta less than 1~GeV/$c$  and
at least 3~mm for candidates with momenta greater than
1~GeV/$c$.  We make a momentum-dependent cut on
the $\pi^+\pi^-$ invariant mass.

Isolated showers with energies greater than
$30$~MeV in the central region of the CsI detector, $|\cos\theta| < 0.71$,
where $\theta$ is the angle with respect to the beam axis,
and greater than $50$~MeV elsewhere, are defined to be photons.
Pairs of photons with an invariant mass within two standard
deviations of the nominal $\pi^0$ mass \cite{pdb94}
are kinematically fitted with the mass constrained to the
$\pi^0$ mass.  To reduce combinatoric backgrounds we
require that the $\pi^0$ momentum be greater than $175$~MeV/$c$, that the
lateral shapes of the showers be consistent with those from photons, and that
$|\cos\theta^*|<0.97$, where $\theta^*$ is the angle between the
direction of flight of the $\pi^0$ and the photons in the $\pi^0$ rest
frame.

We form $\rho$ candidates from
$\pi^+\pi^-$ or $\pi^+\pi^0$ pairs with an invariant mass within $150$~MeV
of the nominal $\rho$ masses.
$K^*$ candidates are selected from $K^+ \pi^-$, $K^+\pi^0$,
$K_S^0\pi^+$ or $K_S^0\pi^0$ pairs \cite{kstars} with
an invariant mass within $75$~MeV of the nominal
$K^*$ masses.  We form $\phi$ candidates from $K^+K^-$ pairs with
invariant mass within $6.5$~MeV  of the nominal $\phi$ mass.

\begin{figure}
\centerline{\hbox{\psfig{figure=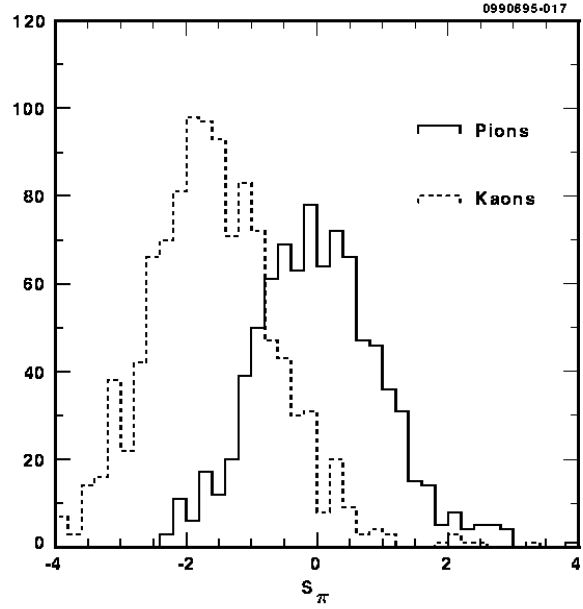,height=6.0in}}}
\vskip 0.5in
\caption{Distribution of $S_\pi$ for kinematically identified high
momentum kaons and pions from $D^{*+} \rightarrow D^0\pi^+;\
D^0\rightarrow K^-\pi^+$ decays.  The solid line shows $S_\pi$
for pions and the dashed line shows $S_\pi$ for kaons.}
\label{fig:dedx}
\end{figure}

Charged particles are identified
as kaons or pions according to $dE/dx$.
We first reject  electrons
based on $dE/dx$ and the ratio of the track momentum to the associated shower
energy in the CsI calorimeter.
We reject muons
by requiring that the tracks not penetrate the steel absorber to a
depth of five nuclear interaction lengths.
We define $S$ for a particular hadron hypothesis as
\begin{equation}
S_{\rm hypothesis} =
{{dE \over dx}|_{\rm measured} - {dE \over dx}|_{\rm hypothesis} \over \sigma}
\end{equation}
where $\sigma$ is the expected resolution, which depends
primarily on the number of
hits used in the $dE/dx$ measurement.
We measure the $S$ distribution in data for kaons and pions using
$D^0\rightarrow K^-\pi^+$ decays where the $D^0$ flavor is tagged
using $D^{*+}\rightarrow D^0\pi^+$ decays.  In particular, we are
interested in separating  pions and kaons with momenta near 2.6 GeV/$c$.
The $S_\pi$ distribution for
the pion hypothesis is shown in Figure~\ref{fig:dedx} for pions and
kaons with momenta
between 2.3 and 3.0~GeV/$c$.  At these momenta, pions and kaons are
separated by  $1.8\pm0.1$ in $S_\pi$.

\section{Candidate selection}
\subsection{Energy Constraint}
Since the $B$'s are
produced via $e^+e^-\rightarrow
\Upsilon (4S)\rightarrow B\bar B$, where the $\Upsilon (4S)$ is at rest in
the lab frame, the energy of either of the two $B$'s is given by the
beam energy, $E_{\rm b}$.  We define
$\Delta E = E_1 + E_2 - E_{\rm b}$ where $E_1$ and $E_2$
are the energies of the daughters of the $B$ meson candidate.
The $\Delta E$ distribution for signal peaks at $\Delta E =0$,
while the background distribution falls linearly
in $\Delta E$ over the region of interest.
The resolution of $\Delta E$ is mode dependent
and in some cases helicity angle dependent (see section III.C)
because of the difference in energy resolution between
neutral and charged pions.
For modes including high momentum
neutral pions in the final state, the $\Delta E$
resolution tends to be asymmetric because of energy loss out of the
back of the CsI crystals.
The $\Delta E$ resolutions for the modes in this paper,
obtained from Monte Carlo simulation, are listed in Tables~\ref{tab:deltae}
and \ref{tab:mlfits}.

We check that the Monte Carlo accurately reproduces the data in
two ways.  First,
the r.m.s.\ $\Delta E$ resolution for $B^0\rightarrow h^+h^-$
(where $h^\pm$ indicates a $\pi^\pm$ or $K^\pm$) is given by
$\sigma_{{\Delta E}_{h^+h^-}} = \sqrt{2}\sigma_p$ where $\sigma_p$ is
the r.m.s.\ momentum resolution at $p=2.6$\ GeV$/c$.
We measure the momentum resolution at $p=5.3$ GeV$/c$ using muon
pairs and in the range $p=1.5$--2.5 GeV$/c$ using the modes
$B \rightarrow \psi K$,
$B\rightarrow D\pi$, and $B \rightarrow D^*\pi$.
We find
$\sigma_{\Delta E_{h^+h^-}} = 24.7\pm 2.3 ^{+1.4}_{-0.7}$ MeV,
where the first error is statistical and the second is
systematic.  This result is in good agreement with the Monte Carlo prediction.
We also test our Monte Carlo simulation
in the modes $B^+ \rightarrow \bar D^0\pi^+$ and $B^0\rightarrow D^-\pi^+$
(where
$\bar D^0 \rightarrow K^+\pi^-$, $\bar D^0 \rightarrow K_S^0\pi^0$, and
$D^-\rightarrow K_S^0\pi^-$)
using an analysis similar to our $B \rightarrow K^*\pi$
analysis.  Again, $\Delta E$ resolutions for data and Monte Carlo are
in good agreement.

The energy constraint also
helps to distinguish between modes of
the same topology.  When a real $K$ is reconstructed as a
$\pi$, $\Delta E$ will peak below zero by an amount dependent on
the particle's momentum.  For example, $\Delta E$
for  $B\rightarrow K^+\pi^-$,  calculated assuming $B\rightarrow\pi^+\pi^-$,
has a distribution which is centered at $-42$~MeV, giving
a separation
of $1.7\sigma$ between $B \rightarrow K^+\pi^-$ and
$B \rightarrow \pi^+\pi^-$.

\subsection{Beam-Constrained Mass}
Since the energy of a $B$ meson is
equal to the beam energy, we use
$E_{\rm b}$\ instead of the reconstructed energy of
the $B$ candidate to calculate  the beam-constrained $B$ mass:
$M_B = \sqrt{E_{\rm b}^2 - {\bf p}_B^2}$.
The beam constraint improves the mass resolution by
about an order of magnitude, since $|{\bf p}_B|$ is only $0.3~$GeV/$c$
and the beam energy is known to
much higher precision than the measured energy of the $B$ decay products.
Mass resolutions range from 2.5 to 3.0 MeV,  where the larger
resolution corresponds to decay modes with high momentum $\pi^0$'s.
Again, we verify the accuracy of our Monte Carlo by studying
fully reconstructed $B$ decays.

\begin{figure}
\centerline{\hbox{\psfig{figure=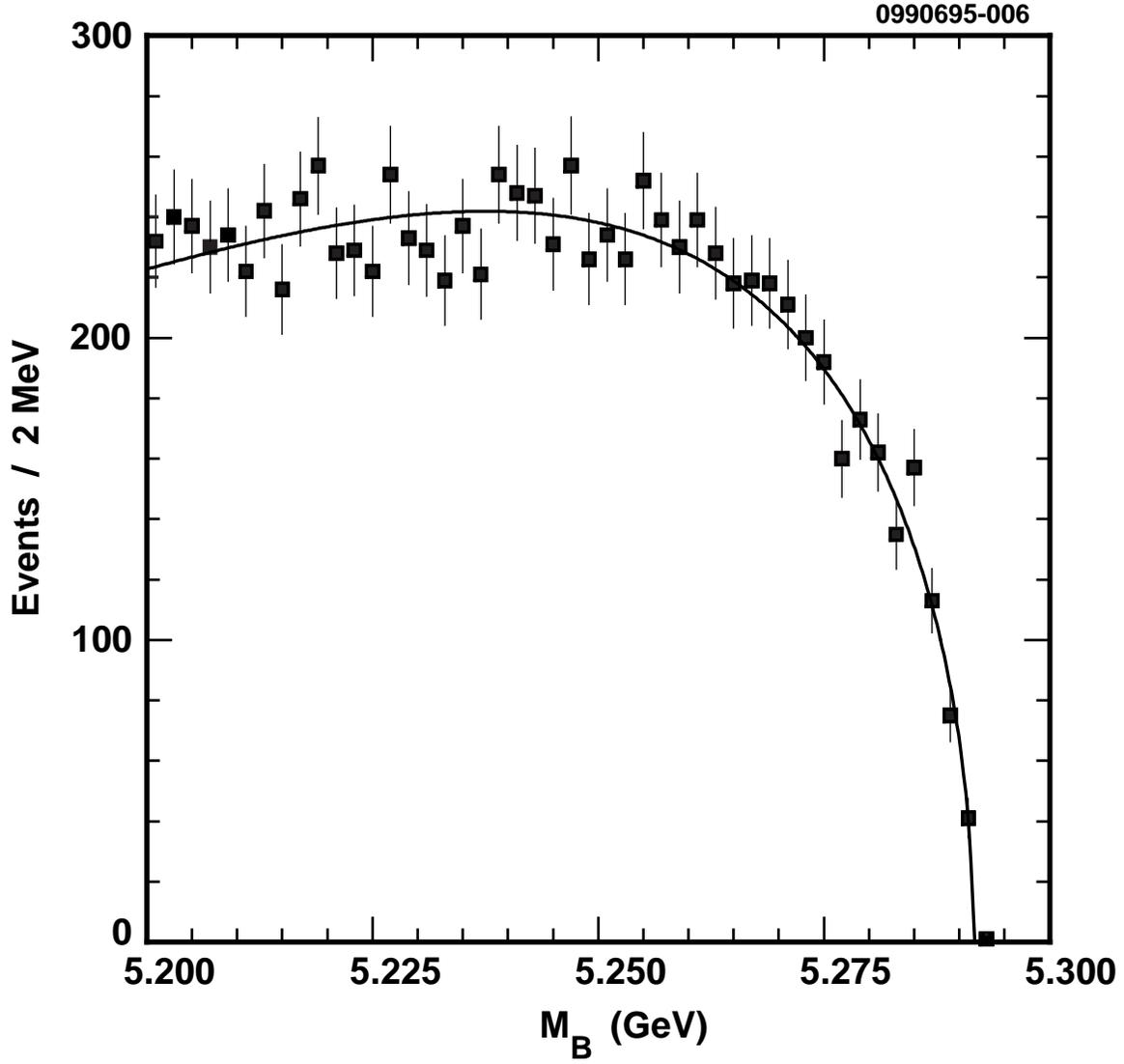,height=6.0in}}}
\vskip .5in
\caption{$M_B$ distribution from below-threshold
background events (squares) and the fit to the
parameterization given in the text (curve).
The mass for the below-threshold data is shifted up to match
the kinematic endpoint of the on-resonance data.}
\label{fig:argusfcn}
\end{figure}

The $M_B$ distribution for continuum background is described by the
empirical shape
\begin{equation}
f(M_B) \propto M_B\sqrt{1-x^2}\exp\left(-\xi(1-x^2)\right)
\end{equation}
where  $x$ is defined as $M_B/E_{\rm b}$ and $\xi$ is a parameter to be fit.
As an example,
Figure~\ref{fig:argusfcn} shows the fit for $B \rightarrow h^+\pi^0$
background from
data taken below $B\bar B$ threshold.

\subsection{Helicity Angle}
The decays $B \rightarrow \pi\rho$,
$B \rightarrow K\rho$, $B \rightarrow K^*\pi$,
and $B \rightarrow K\phi$ are of the form
pseudoscalar $\rightarrow$ vector + pseudoscalar.  Therefore
we expect the helicity angle, $\theta_H$, between a resonance
daughter direction and the $B$ direction in the resonance rest
frame to have a $\cos^2\theta_H$ distribution.  For these decays
we require $|\cos \theta_H| > 0.5$.

\subsection{$D$ Veto}
We suppress
events from the decay $B^+\rightarrow \bar{D}^0 \pi^+$
(where $\bar{D}^0 \rightarrow K^+\pi^-$ or $\bar{D}^0 \rightarrow
K_S^0\pi^0$) or $B^0 \rightarrow D^-\pi^+$ (where $D^-\rightarrow
K_S^0\pi^-$) by rejecting any candidate that can be
interpreted as $B\rightarrow \bar{D} \pi$, with a $K\pi$
invariant mass within $2\sigma$ of the nominal $D$
mass.  We expect less than half an event background per mode from
$B\rightarrow \bar D \pi$ events after this veto.
The vetoed $D\pi$ signal is used
as a cross-check of signal distributions and efficiencies.

\section{Background Suppression using Event Shape}

The dominant background in all modes is from continuum production, $e^+e^-
\rightarrow q\bar{q}\ (q=u,d,s,c)$.  After the $D$ veto,
background from $b \rightarrow c$
decays is negligible
in all modes because final state particles in such
decays have maximum momenta lower than what is required for the decays of
interest here.  We have also studied backgrounds from the
rare processes
$b \rightarrow s \gamma$ and $b \rightarrow u\ell\nu$ and find
these to be negligible as well.

Since the $B$ mesons are approximately at rest in the lab,
the angles of the decay products of the two $B$ decays
are uncorrelated
and the event looks spherical.  On the other hand,
hadrons from continuum $q\bar{q}$ production
tend to display a two-jet structure.
This event shape distinction is exploited in two ways.

\begin{figure}
\centerline{\hbox{\psfig{figure=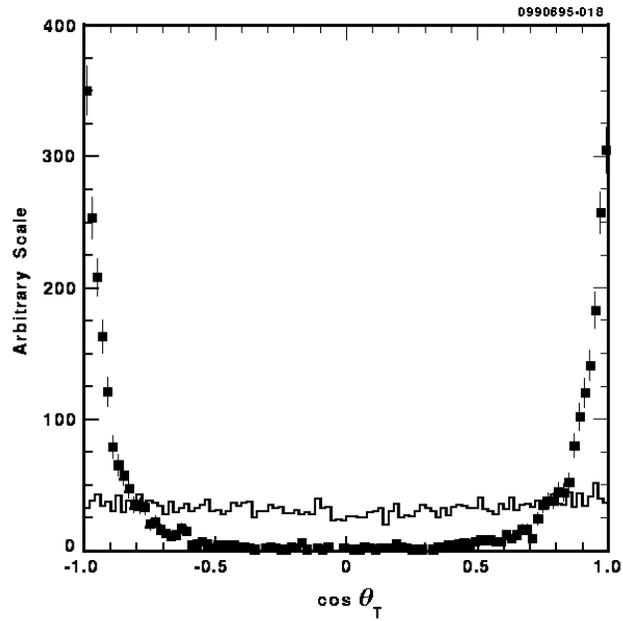,height=6.0in}}}
\vskip .5in
\caption{The $\cos \theta_T$ distributions for
background data (squares)
and $B^0 \rightarrow \pi^+\pi^-$ Monte Carlo signal (histogram).}
\label{fig:costhrcomp}
\end{figure}

First, we calculate the angle, $\theta_T$,
between the thrust axis of the $B$ candidate and the thrust
axis of all the remaining charged and neutral particles in the event.
The distribution of $\cos \theta_T$ is strongly peaked
near $\pm1$ for $q\bar q$ events and is nearly flat for $B\bar B$ events.
Figure~\ref{fig:costhrcomp} compares the $\cos \theta_T$ distributions
for Monte Carlo signal events and background data.
We require $|\cos\theta_T| < 0.7$ which
removes more than $90\%$ of the continuum background with
approximately $65\%$ efficiency for signal events \cite{thrustnotflat}.

\begin{figure}
\centerline{\hbox{\psfig{figure=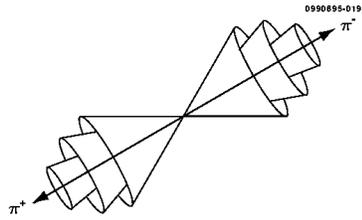,height=4in}}}
\vskip .5in
\caption{Illustration of the first three of the nine polar angle
intervals.}
\label{fig:vcal}
\end{figure}

Second, we characterize the event shape by
dividing the space around the candidate thrust axis into nine
polar angle intervals
of $10^\circ$ each, illustrated in Figure~\ref{fig:vcal};
the $i^{th}$ interval covers angles
with respect to the candidate thrust axis
from $(i-1)\times 10^{\circ}$\ to $i\times 10^{\circ}$.  We fold
the event such that the forward and backward intervals are combined.
We then define the momentum flow, $x_i$ ($i=1,9$), into the
$i^{th}$ interval as the scalar sum of the momenta of all charged tracks and
neutral showers pointing in that interval.
The $10^\circ$ binning was chosen to enhance the
distinction between $B \bar B$
and continuum background events.

Angular momentum conservation considerations provide additional distinction
between $B\bar B$ and continuum $q\bar q$ events.
In $q\bar q$ events, the direction of the candidate thrust axis,
$\theta_{q\bar q}$, with respect to the beam axis in the lab frame
tends to maintain the
$1+\cos^2 \theta_{q\bar q}$ distribution of the primary
quarks.  The direction of the candidate thrust
axis for $B \bar B$ events is random.
The candidate $B$ direction, $\theta_B$, with respect to
the beam axis exhibits a $\sin^2 \theta_B$ distribution for
$B \bar B$ events and is random for $q \bar q$ events.

A Fisher discriminant\cite{kendall}
is formed from these eleven
variables: the nine
momentum flow variables, $|\cos \theta_{q\bar q}|$,
and $|\cos \theta_B|$.
The discriminant, $\cal F$, is the linear combination
\begin{equation}
{\cal F} = \sum^{11}_{i=1} \alpha_i\ x_i
\label{eqn:xfdef}
\end{equation}
of the input variables, $x_i$, that maximizes the separation between
signal and background.
The Fisher discriminant parameters, $\alpha_i$, are given by
\begin{equation}
\alpha_i = \sum_{j=1}^{11} (U_{ij}^b + U_{ij}^{s})^{-1} \times
(\mu_j^{b} - \mu_j^{s}).
\label{eqn:alphadef}
\end{equation}
where $U_{ij}^s$\  and $U_{ij}^{b}$\ are the covariance matrices of the
input variables for
signal and background events, and $\mu_j^s,\ \mu_j^{b}$\
are the mean values of the input variables.
We calculate $\alpha_i$ using Monte Carlo samples of
signal and background events
in the mode
$B \rightarrow \pi^+\pi^-$.

\begin{figure}
\centerline{\hbox{\psfig{figure=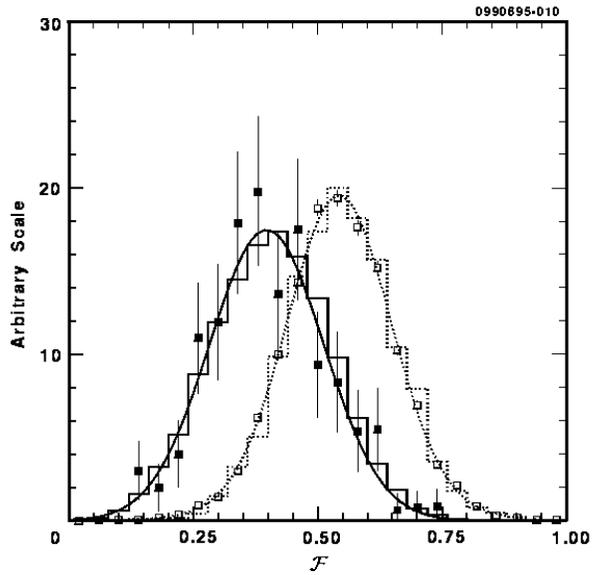,height=6in}}}
\vskip .5in
\caption{The $\cal F$ distribution for $B^0\rightarrow \pi^+\pi^-$
Monte Carlo (solid histogram), $B \rightarrow D\pi$ signal data
(filled squares), the fit to the signal data (solid curve),
the background Monte Carlo (dotted histogram), background
data (open squares), and the fit to the background data
(dotted curve).}
\label{fig:fdstan}
\end{figure}

\begin{figure}
\centerline{\hbox{\psfig{figure=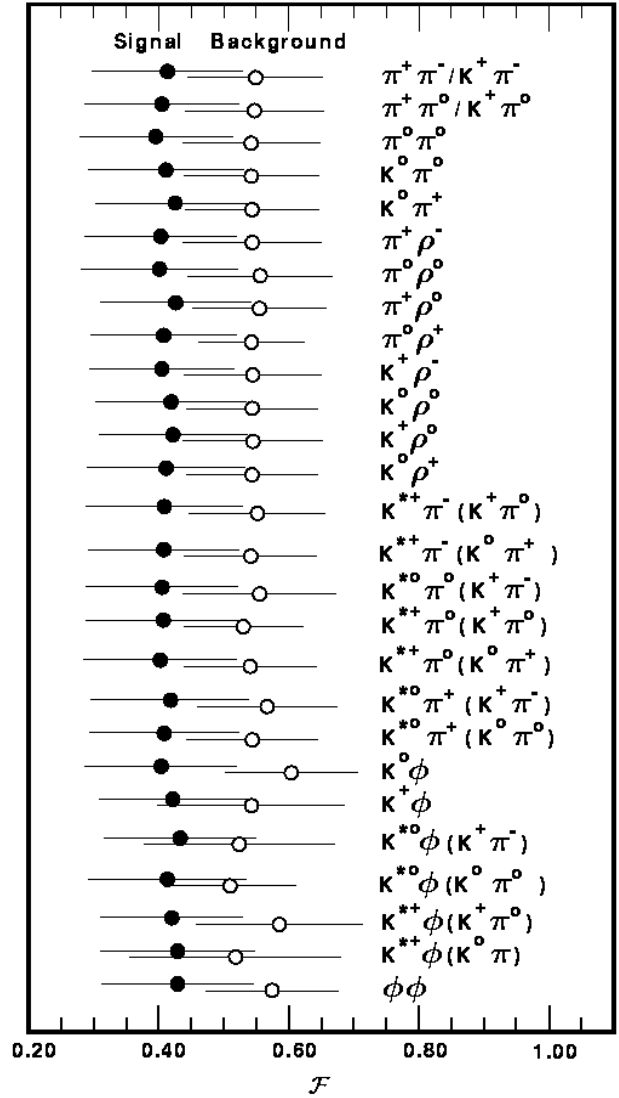,height=7in}}}
\vskip .5in
\caption{The means of the Fisher output distributions for
signal (filled circles) and background (open circles) for
the modes in this study.
The error bars indicate the width of the distributions.  Since
the backgrounds in the $\phi$ modes are small, their background
means are poorly measured.}
\label{fig:fdmiracle}
\end{figure}

Figure~\ref{fig:fdstan} shows the $\cal F$ distributions for Monte Carlo
signal in the mode $B^0\rightarrow \pi^+\pi^-$, and data signal in
the modes $B \rightarrow \bar D\pi$.  Figure~\ref{fig:fdstan} also shows
the $\cal F$ distributions for Monte Carlo background in the mode
$B\rightarrow h^+\pi^-$ and
below-threshold background
data for modes comprising three charged tracks
or two charged tracks and a $\pi^0$.
The $\cal F$
distribution for signal is fit by a Gaussian distribution,
while the $\cal F$ distribution for background data is best fit
by the sum of two Gaussians with the same mean but different variances and
normalizations.
The separation between signal and background means is approximately 1.3 times
the signal width.
We find that the Fisher coefficients calculated for $B^0 \rightarrow
\pi^+\pi^-$ work equally well for all other decay modes presented in this
paper.
Figure~\ref{fig:fdmiracle} shows the remarkable consistency of the
means and widths of the $\cal F$
distributions for signal and background Monte Carlo for the modes in
this study.

\section{Analysis}
For the decay modes $B \rightarrow \pi\pi$, $B \rightarrow K\pi$,
and $B \rightarrow KK$,
we extract the signal yield using a maximum likelihood fit. For the other
decay modes, we use a simple counting analysis.  Both techniques
are described below.

\subsection {Maximum Likelihood Fit}
We perform unbinned
maximum likelihood fits using $\Delta E$, $M_B$, $\cal F$,
and $dE/dx$ (where appropriate) as input information for each candidate
event to determine the signal yields for
$B^0\rightarrow \pi^+\pi^-,\ K^+\pi^-,\ K^+K^-,\
\pi^0\pi^0,\ K^0\pi^0$, and
$B^+\rightarrow \pi^+\pi^0,\ K^+\pi^0,\
K^0\pi^+$.
Five different fits are performed as listed in
Table~\ref{tab:mlfits}.

For each fit a likelihood function $\cal L$ is defined as:
\begin{equation}
{\cal L} =
\prod_{i=1}^{N} P\left( f_1, ..., f_m;
\left(\Delta E, M_B, {\cal F}, dE/dx\right)_i\right)
\label{eqn:likefun}
\end{equation}
where $P\left( f_1, ..., f_m;
\left(\Delta E, M_B, {\cal F}, dE/dx\right)_i\right)$
is the probability density function
evaluated at the measured point
$(\Delta E,\ M_B,\ {\cal F},\ dE/dx)_i$ for a
single candidate event, $i$,
for some assumption of the values
of the yield fractions, $f_j$, that are determined
by the fit.  $N$
is the total number of events that are fit.
The fit includes all the candidate events
that pass the selection criteria discussed above as well as
$|\cos\theta_T|<0.7$, and $0<{\cal F}<1$.
The $\Delta E$ and $M_B$ fit ranges are given in Table~\ref{tab:mlfits}.

For the case of $B\rightarrow h^+h^-$, the
probability $P_i = P\left( f_1, ..., f_m;
\left(\Delta E, M_B, {\cal F}, dE/dx\right)_i\right)$
is then defined by:
\begin{eqnarray}
P_i &= & f^S_{\pi\pi} P^S_{\pi\pi} + f^S_{K\pi} P^S_{K\pi} +
f^S_{KK} P^S_{KK}
% \nonumber \\ &&
+ (1- f^S_{\pi\pi} - f^S_{K\pi} - f^S_{KK}) P^C \\
P^C &= & f^C_{\pi\pi} P^C_{\pi\pi} + f^C_{K\pi} P^C_{K\pi}
+ (1- f^C_{\pi\pi} - f^C_{K\pi}) P^C_{KK} \nonumber
\end{eqnarray}
where, for example, $P^S_{\pi\pi}$ ($P^C_{\pi\pi}$)
is the product of the individual
probability density functions for $\Delta E$, $M_B$, $\cal F$, and
$dE/dx$ for $\pi^+\pi^-$
signal (continuum background).  The signal yield in
$B^0 \rightarrow \pi^+\pi^-$, for example, is then given by
$N_{\pi\pi} \equiv f_{\pi\pi}^S\times N$.

\begin{figure}
\centerline{\hbox{\psfig{figure=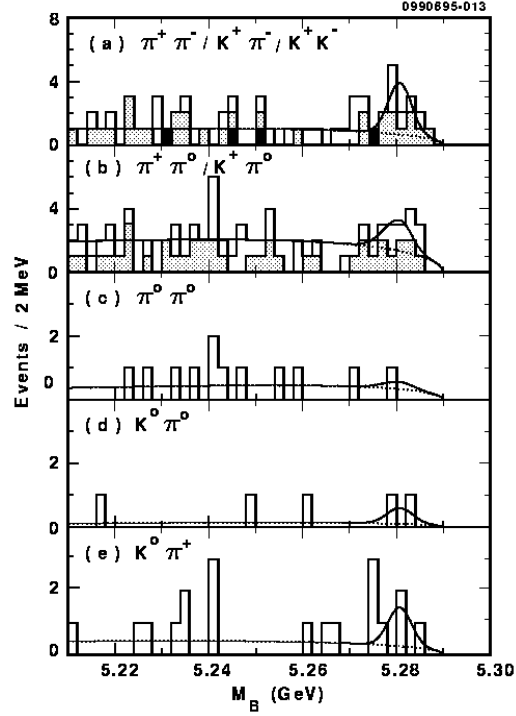,height=6in}}}
\vskip 1in
\caption{$M_B$ plots for (a) $B^0\rightarrow \pi^+\pi^-$ (unshaded)
$B^0\rightarrow K^+\pi^-$ (grey), and $B^0\rightarrow K^+K^-$, (black)
(b) $B^+\rightarrow \pi^+\pi^0$ (unshaded) and $B^+\rightarrow
K^+\pi^0$ (grey), (c) $B^0\rightarrow \pi^0\pi^0$, (d)
$B^0 \rightarrow K^0\pi^0$, and e) $B\rightarrow K^0\pi^+$.
The projection of the total likelihood fit (solid curve)
and the continuum background component (dotted curve) are overlaid.}
\label{fig:mass_2body}
\end{figure}

\begin{figure}
\centerline{\hbox{\psfig{figure=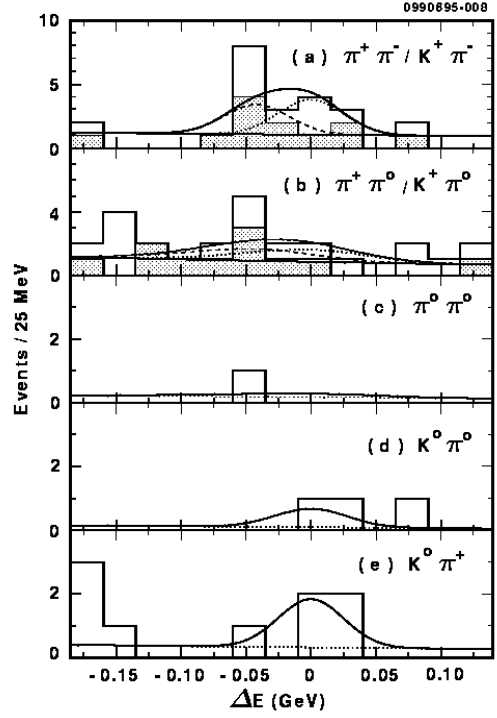,height=6in}}}
\vskip 1in
\caption{$\Delta E$ plots for (a) $B^0\rightarrow \pi^+\pi^-$ (unshaded)
$B^0\rightarrow K^+\pi^-$ (grey),
(b) $B^+\rightarrow \pi^+\pi^0$ (unshaded) and $B^+\rightarrow
K^+\pi^0$ (grey), (c) $B^0\rightarrow \pi^0\pi^0$, (d)
$B^0 \rightarrow K^0\pi^0$, and (e) $B\rightarrow K^0\pi^+$.
In plots (a) and (b),
the projection of the total likelihood fit (upper solid curve),
the $\pi\pi$ signal component (dotted curve), the
$K\pi$ signal component (dashed curve), and the background
component (lower solid curve) are overlaid.  In plots
(c)-(e), the total likelihood fit (solid curve) and
background component (dotted curve) are overlaid.}
\label{fig:de_2body}
\end{figure}

The central values of the individual signal yields
from the fits are given in
Table~\ref{tab:mlresults}.  None of the individual modes shows a
statistically compelling signal.  To illustrate the fits,
Figure~\ref{fig:mass_2body} shows $M_B$ projections for events
in a signal region defined by
$|\Delta E| < 2\sigma_{\Delta E}$ and ${\cal F} < 0.5$
and Figure~\ref{fig:de_2body} shows the $\Delta E$ projections
for events within a 2$\sigma$ $M_B$ cut and
${\cal F} < 0.5$.
The modes are sorted by $dE/dx$ according to the most likely
hypothesis and are shown in the plots with different shadings.
Overlaid on these plots are the projections of the
fit function integrated over the remaining variables
within these cuts.  (Note that these curves are not fits to these
particular histograms.)

\begin{figure}
\centerline{\hbox{\psfig{figure=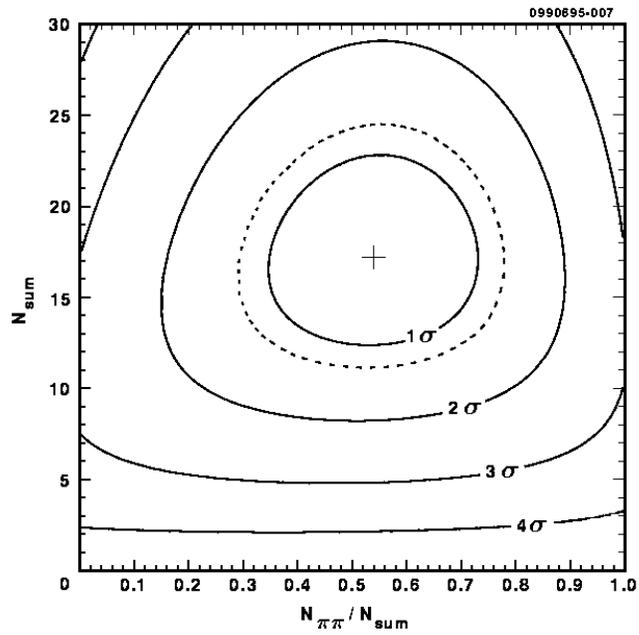,height=6in}}}
\vskip 0.5in
\caption{The central value ($+$) of the likelihood fit to
$N_{\rm sum} \equiv N_{\pi\pi} + N_{K\pi}$ and
$R \equiv  N_{\pi\pi}/N_{\rm sum}$ for
$B^0\rightarrow \pi^+\pi^-$
and $B^0\rightarrow K^+\pi^-$.  The solid curves are the
$n\sigma$ contours, and the dotted curve is the 1.28$\sigma$
contour.}
\label{fig:contour}
\end{figure}

Our previous publication \cite{kpi} reported a significant signal in the
sum of $B^0 \rightarrow \pi^+\pi^-$ and $B^0 \rightarrow K^+\pi^-$.
While our current analysis confirms this result, we now focus on
separating the two modes.  We separate
the systematic errors
that affect the total yield from those that affect the
separation of the two modes.
We do this by  repeating the likelihood fit using
$N_{\rm sum}\equiv N_{\pi\pi} + N_{K\pi}$,
$R \equiv N_{\pi\pi}/N_{\rm sum}$, and fixing $N_{KK} = 0$, its most
likely value.
We find:
\begin{eqnarray*}
 N_{\rm sum}
= && 17.2^{+5.6\ +2.2}_{-4.9\ -2.5} \nonumber \\
 R = &&0.54^{+0.19}_{-0.20}\pm 0.05
\end{eqnarray*}
where the first error is statistical and the second is
systematic (described below).
The result of this fit is shown in Figure~\ref{fig:contour}.
This figure shows a contour plot (statistical errors only) of
$N_{\rm sum}$ {\it vs.}\ $R$ in which the solid curves represent the
$n\sigma$ contours ($n=$1--4) corresponding to decreases in the
log likelihood by $0.5n^2$.  The dashed curve represents the
$1.28\sigma$ contour, from which estimates of the 90\%
confidence level limits
can be obtained.
The central value of $N_{\rm sum}$ has a statistical significance of
$5.2\sigma$.  The significance is reduced to
$4.2\sigma$\ if all parameters defining $\cal L$\
are varied coherently so as to minimize $N_{\rm sum}$.
Further support for the statistical significance of
the result is obtained by using Monte Carlo to draw 10000
sample experiments, each
with the same number of events as in the
data fit region but no signal events.
We then fit each of these sample experiments to determine
$N_{\rm sum}$\ in the same way as done for data.
We find that none of the 10000 sample experiments leads to
$N_{\rm sum}>10$.

None of the physical range of $R$\ can be excluded
at the $3\sigma$\ level.  However the systematic error
of $R$ is only 10\% (see below and Table~\ref{tab:mlsyst}).
We therefore conclude that our analysis technique has sufficient
power to distinguish the $\pi^+\pi^-$
mode from $K^+\pi^-$, but at this time
we do not have the statistics to do so.

Since none of our fits has a statistically significant signal, we
calculate the $90\%$ confidence level upper limit yield
from the fit, $N^{90}$, given by
\begin{equation}
{\int_0^{N^{90}} {\cal L}_{\rm max} (N) dN
\over
\int_0^{\infty} {\cal L}_{\rm max} (N) dN}
= 0.90
\label{eqn:upplim}
\end{equation}
where ${\cal L}_{\rm max}(N)$ is
the maximal $\cal L$\ at fixed $N$\ to
conservatively account for possible correlations among
the free parameters in the fit.
The upper limit yield is then increased by the systematic error
determined by varying the parameters defining
$\cal L$\ within their systematic uncertainty as discussed below.
Table~\ref{tab:mlresults} summarizes upper limits on the yields for
the various decay modes.

To determine the systematic effects on the yield due
to uncertainty of the shapes used in the likelihood fits,
we vary the parameters
that define the likelihood functions.
The variations of the yields are given in  Table~\ref{tab:mlsyst}.
The largest contribution
to the systematic error arises from the variation of the $M_B$
background shape.
For this
shape, $f(M_B) \propto M_B\sqrt{1-x^2}\exp(-\xi(1-x^2))$
($x \equiv M_B/E_{\rm b}$), we vary $E_{\rm b}$ by $\pm1$ MeV,
consistent with
observed variation; we vary $\xi$ by the amount allowed by a fit
to background data (below-threshold and on-resonance
$\Delta E$ sideband) which pass all other selection criteria.  To be
conservative, we allow for correlated variations of $E_{\rm b}$ and
$\xi$.

\subsection {Event-Counting Analyses}
In the event-counting analyses we make cuts on
$\Delta E$, $M_B$, $\cal F$, and $dE/dx$.
The cuts for $\Delta E$ and $M_B$
are mode dependent and are listed in Table~\ref{tab:deltae}.
We require ${\cal F} < 0.5$.  Tracks are
identified as kaons and/or pions if their specific ionization loss,
$dE/dx$, is within three standard deviations of the expected value.
For certain topologies, candidates can have multiple
interpretations under different particle hypotheses.  In these cases
we use a strict identification scheme where a track is positively
identified as a kaon or a pion depending on which $dE/dx$ hypothesis
is more likely:
we sort the modes with two charged tracks plus a $\pi^0$
($\pi^+\rho^-$, $\pi^0\rho^0$, $K^+\rho^-$,
$K^{*+}\pi^-$, and $K^{*0}\pi^0$)
by requiring strict identification for both charged tracks.
For modes with three charged tracks
($\pi^+\rho^0$, $K^+\rho^0$, and $K^{*0}\pi^+$)
we require strict identification of the two like-sign tracks,
while the unlike-sign track \cite{chargestrangeness} is
required to be consistent with the pion hypothesis within two
standard deviations.
We separate modes with one charged track plus two $\pi^0$'s
($\rho^+\pi^0$ and $K^{*+}\pi^0$)
by requiring strict identification of the charged track.

\begin{figure}
\centerline{\hbox{\psfig{figure=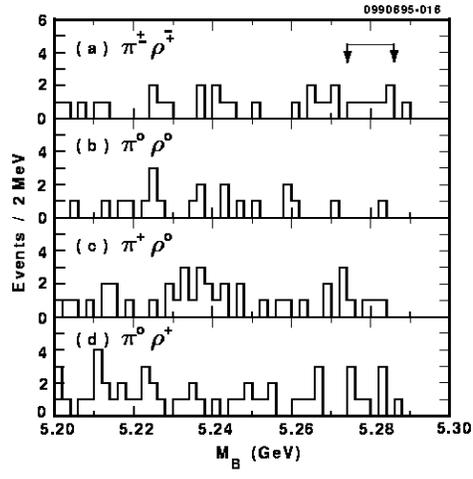,height=5.5in}}}
\vskip 1.0in
\caption{$M_B$ plots for (a) $B^0\rightarrow \pi^\pm\rho^\mp$,
(b) $B^0\rightarrow \pi^0\rho^0$, (c) $B^+ \rightarrow \pi^+\rho^0$, and
(d) $B^+\rightarrow \pi^0\rho^+$.  The signal region is indicated.}
\label{fig:mass_pirho}
\end{figure}

\begin{figure}
\centerline{\hbox{\psfig{figure=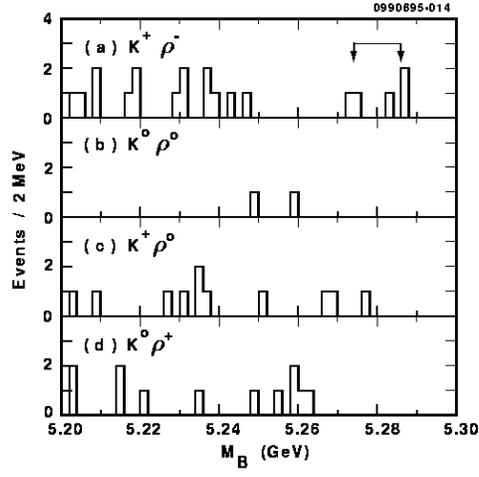,height=5.5in}}}
\vskip 1.0in
\caption{$M_B$ plots for (a) $B^0\rightarrow K^+\rho^-$,
(b) $B^0\rightarrow K^0\rho^0$, (c) $B^+ \rightarrow K^+\rho^0$, and
(d) $B^+\rightarrow K^0\rho^+$.  The signal region is indicated.}
\label{fig:mass_krho}
\end{figure}

\begin{figure}
\centerline{\hbox{\psfig{figure=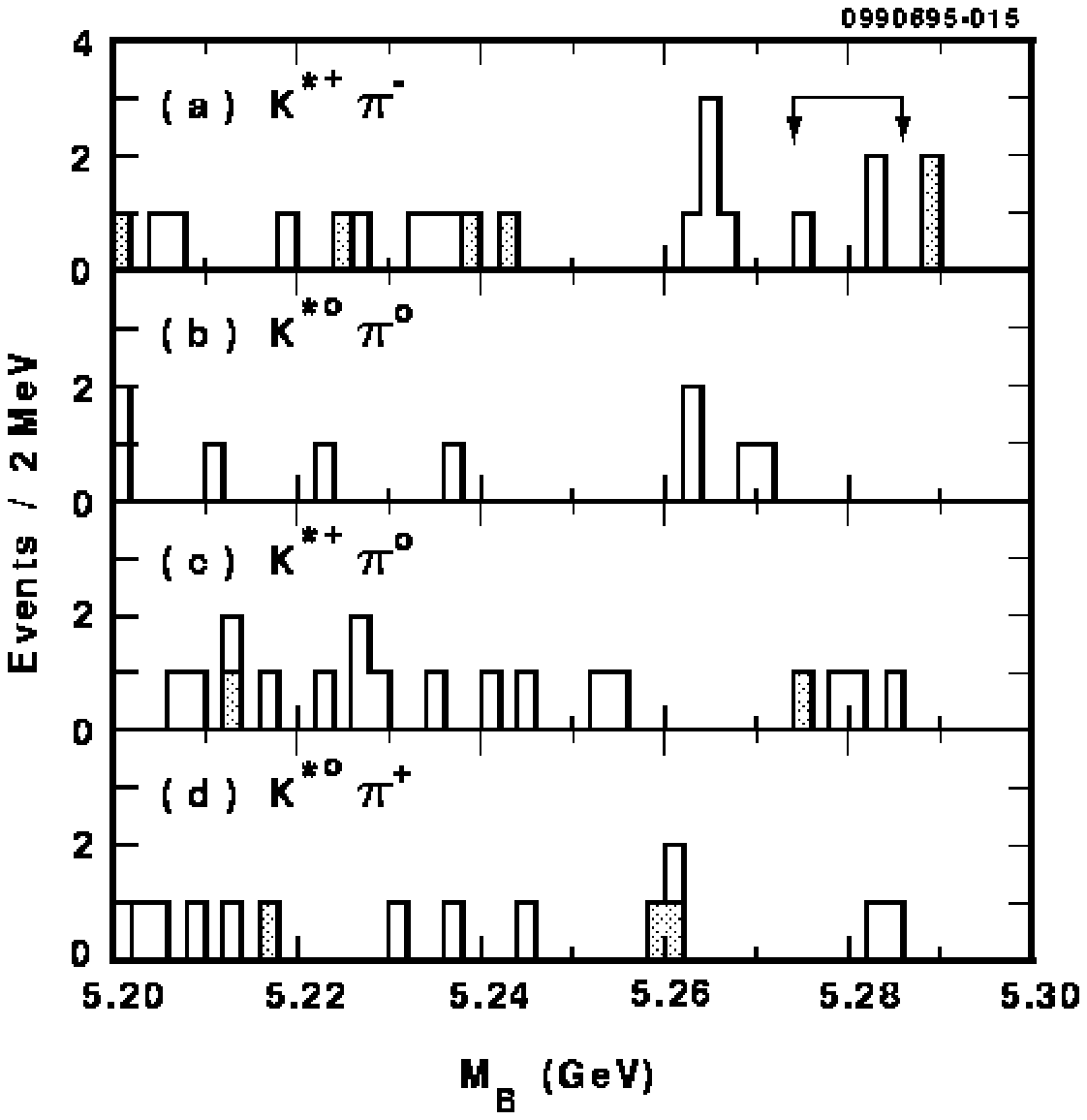,height=5.5in}}}
\vskip 1.0in
\caption{$M_B$ plots for (a) $B^0\rightarrow K^{*+}\pi^-$,
(b) $B^0\rightarrow K^{*0}\pi^0$, (c) $B^+ \rightarrow K^{*+}\pi^0$, and
(d) $B^+\rightarrow K^{*0}\pi^+$. The shaded events are from
$K^* \rightarrow K_S^0\pi$ decay modes and the unshaded events are from
$K^* \rightarrow K^+\pi$ decay modes.  The signal region is indicated.}
\label{fig:mass_kstarpi}
\end{figure}

\begin{figure}
\centerline{\hbox{\psfig{figure=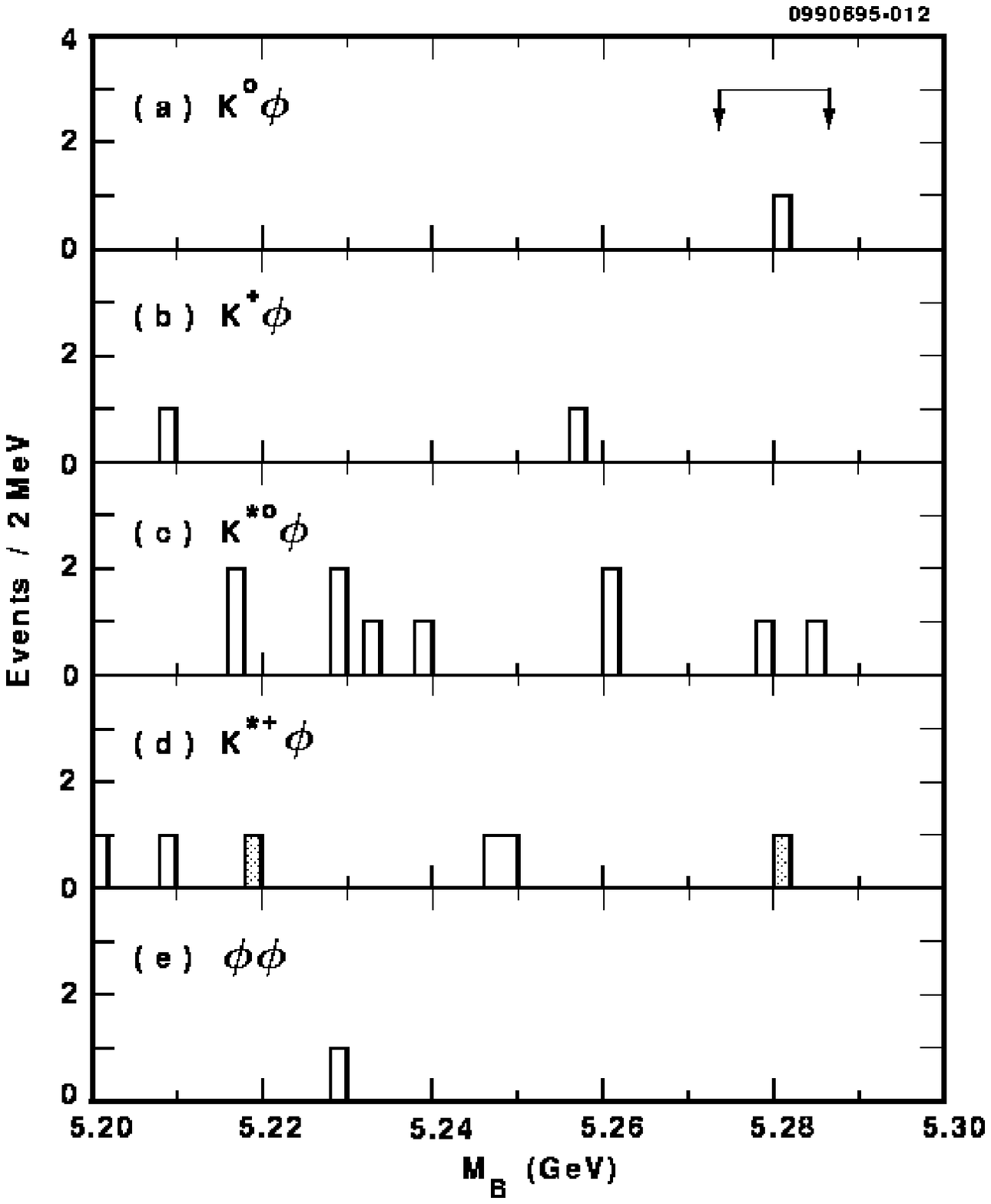,height=5.5in}}}
\vskip 2.5in
\caption{$M_B$ plots for (a) $B^0\rightarrow K^0\phi$,
(b) $B^+\rightarrow K^+\phi$, (c) $B^0 \rightarrow K^{*0}\phi$,
(d) $B^+\rightarrow K^{*+}\phi$, and
(e) $B^0\rightarrow \phi\phi$.  The shaded events are from
$K^*\rightarrow K_S^0\pi$ decay modes and the unshaded events are from
$K^*\rightarrow K^+\pi$ decay modes.  The signal region is indicated.}
\label{fig:mass_kphi}
\end{figure}

Figures~\ref{fig:mass_pirho}--\ref{fig:mass_kphi}
show $M_B$ distributions for $B \rightarrow \pi\rho$,
$B \rightarrow K\rho$, $B\rightarrow K^*\pi$, $B\rightarrow K\phi $,
$B\rightarrow K^*\phi$ and $B\rightarrow \phi\phi$
candidates (after making the cuts on $\Delta E$, $\cal F$, and particle
identification
described above.)  The numbers of events in the signal regions are
listed in Table~\ref{tab:evcount_results}.

In order to estimate the background in our
signal box, we look in a large
sideband region in the $\Delta E~vs.~M_B$ plane:
$5.20 < M_B < 5.27$ GeV and $| \Delta E | < 200$ MeV.
The expected background in the signal region is obtained by
scaling the number of events seen in the
on-resonance  and below-threshold
sideband regions (weighted appropriately for luminosity).
Scale factors are found using a continuum Monte Carlo
sample which is about five times the continuum data on-resonance.
In many modes, the
backgrounds are so low that there are insufficient statistics in the
Monte Carlo to adequately determine a scale factor.  For these modes,
we calculate upper limits assuming all observed events are signal
candidates.  The estimated background for each mode
is also listed in Table~\ref{tab:evcount_results}.

Although we find that there are slight excesses
above expected background
in some modes, no excess is statistically compelling.
We therefore calculate upper limits on the numbers of
signal events using the procedure outlined in the
Review of Particle Properties~\cite{pdb94} for evaluation of upper
limits in the presence of background.
To account for the uncertainties in the estimated continuum
background we reduce the background estimate by its uncertainty
prior to calculating the upper limit on the signal yield.

\section{Efficiencies}
The reconstruction efficiencies were determined using events generated
with a GEANT-based Monte Carlo simulation program~\cite{geant}.
Systematic uncertainties were determined using data wherever possible.
Some of the largest systematic errors
come from uncertainties in the efficiency of the $|\cos\theta_T| < 0.7$
cut (6\%), the uncertainty in
the $\pi^0$ efficiency (7\% per $\pi^0$), and
the uncertainty in the $K_S^0$ efficiency (8\% per $K_S^0$).
In higher
multiplicity modes, substantial contributions come from the
uncertainty in the tracking efficiency (2\% per track).  In the
$B \rightarrow \pi\rho,\ K\rho,\ K^*\pi$ analyses,
the simulation of the efficiency
for the particle identification method has a systematic error of
15\%.  For the event-counting analyses, the uncertainty in the
${\cal F} < 0.5$ cut is 5\%.

The total detection efficiency, ${\cal E}$,
is given by ${\cal E} \equiv {\cal E}_r \times {\cal E}_d$,
where ${\cal E}_r$ is the reconstruction efficiency and
${\cal E}_d$ is the product of the appropriate daughter branching
fractions.  The efficiencies, with systematic errors, are listed in
Tables~\ref{tab:mlresults} and \ref{tab:eff}.

\section{Upper Limit Branching Fractions}
Upper limits on the branching fractions are given by
$N_{\rm UL}/({\cal E} N_B)$
where $N_{\rm UL}$ is the upper limit on the signal yield,
${\cal E}$ is the total detection efficiency, and $N_B$ is the
number of
$B^0$'s or $B^+$'s produced, 2.6 million,  assuming
equal production of charged and
neutral $B$ mesons. To conservatively account for the
systematic uncertainty in our efficiency, we reduce the efficiency by
one standard deviation.  The upper limits on the branching fractions
appear in Tables \ref{tab:mlresults} and \ref{tab:evcount_results}.

\section{Summary and Conclusions}
We have searched for rare hadronic $B$ decays in many modes and
find a signal only in the sum of $\pi^+\pi^-$ and $K^+\pi^-$.
The combined branching fraction, ${\cal B}(\pi^+\pi^- + K^+\pi^-)
= (1.8^{+0.6+0.2}_{-0.5-0.3}\pm0.2) \times 10^{-5}$,
is consistent with our previously published result.
We have presented new upper limits on the branching
fractions for a variety of charmless hadronic decays of $B$~mesons
in the range $10^{-4}$ to $10^{-6}$.
These results are significant improvements over those
previously published.  Our sensitivity is at the level of Standard
Model
predictions for the modes $\pi^+\pi^-,\ K^+\pi^-,\ \pi^+\pi^0,\
K^+\pi^0,\  \pi^{\pm}\rho^{\mp},\ K^+\phi$,\ and $K^{*0}\phi$.

\acknowledgments
We gratefully acknowledge the effort of the
CESR staff in providing us with
excellent luminosity and running conditions.
J.P.A., J.R.P., and I.P.J.S. thank
the NYI program of the NSF,
G.E. thanks the Heisenberg Foundation,
K.K.G., M.S., H.N.N., T.S., and H.Y. thank the
OJI program of DOE,
J.R.P thanks the A.P. Sloan Foundation,
and A.W. thanks the
Alexander von Humboldt Stiftung
for support.
This work was supported by the National Science Foundation, the
U.S. Department of Energy, and the Natural Sciences and
Engineering Research Council of Canada.

\begin{table}
%\mediumtext
\begin{center}
\caption{Resolutions of $\Delta E$ and the signal regions
for $\Delta E$ and $\Delta M_B =  M_B-5280$~MeV
for the event-counting
analyses.  Indicated in parentheses are the $K^*$ decay
modes used.}
\vspace{0.5cm}
\begin{tabular}{lcll}
     &  & \multicolumn{2}{c}{Signal Region} \\
Mode & $\sigma_{\Delta E}$ & $|\Delta E|$ & $|\Delta M_B|$ \\
     & (MeV) & (MeV) & (MeV) \\
\hline
$\pi^\pm\rho^\mp$ & 25--46 & $<2\sigma$\tablenotemark[1]   & $<6.0$  \\
$\pi^0\rho^0$     & 46    & $<90 $      & $<6.0$                     \\
$\pi^+\rho^0$     & 23    & $<50$       & $<6.0$                     \\
$\pi^0\rho^+$     & 50  &  $<100$& $<6.0$                            \\
                  &     &        &         \\
$K^+\rho^-  $     & 25--46 & $<2\sigma$\tablenotemark[1]   & $<6.0$  \\
$K^0\rho^0  $     & 22     & $<50$                         & $<6.0$  \\
$K^+\rho^0  $     & 23    & $<50$       & $<6.0$                     \\
$K^0\rho^+  $     & 22--45 & $<2\sigma$\tablenotemark[1]   & $<6.0$  \\
                  &     &        &           \\

$K^{*+}\pi^-$     &      &                       &       \\
{}~~~~($K^+\pi^0$)&25--40 & $<2\sigma$\tablenotemark[1]   & $<6.0$     \\
{}~~~~($K^0\pi^+$)& 21   & $<50$   & $<6.0$                            \\
$K^{*0}\pi^0$       &           &         &                    \\
{}~~~~($K^+\pi^-$)& 44    & $<90 $                & $<6.0$             \\

$K^{*+}\pi^0$       &      &        &                    \\
{}~~~~($K^+\pi^0$) & 50  & $<100$ & $<6.0$                              \\
{}~~~~($K^0\pi^+$) & 45    & $<90$       & $<6.0$                       \\
$K^{*0}\pi^+$       &      &        &                    \\
{}~~~~($K^+\pi^-$) & 23    & $<50$       & $<6.0$                       \\
{}~~~~($K^0\pi^0$) & 22--40 & $<2\sigma$\tablenotemark[1]    & $<6.0$   \\
                  &     &        &           \\

$K^0\phi$	  & 18    & $<45$       & $<6.5$    \\
$K^+\phi$	  & 23    & $<60$       & $<6.5$    \\
$K^{*0}\phi$        &            &         &   \\
{}~~~~~~$(K^+\pi^-)$  & 20    & $<50$       & $<6.5$  \\
{}~~~~~~$(K^0\pi^0)$  & 24    & $<60$       & $<6.5$  \\
$K^{*+}\phi$     &        &         &       \\
{}~~~~~~$(K^+\pi^0)$  & 23    & $<60$       & $<6.5$  \\
{}~~~~~~$(K^0\pi^+)$  & 17    & $<45$       & $<6.5$  \\
$\phi\phi$	  & 16    & $<40$         & $<6.5$  \\
\end{tabular}
\tablenotemark[1]{The $\Delta E$ resolution and cut are functions
of the helicity angle.}
\label{tab:deltae}
\end{center}
\end{table}
\clearpage

\begin{table}
\mediumtext
\begin{center}
\caption{Resolutions of $\Delta E$, the fit regions in $\Delta E$ and
$M_B$, and the number of events, $N$, in the fit regions for the
likelihood analyses.}
\begin{tabular}{l c c c c}
     &                    &\multicolumn{2}{c}{Fit region} & \\
 & $\sigma_{\Delta E}$ &   $\Delta E$ & $M_B$         &  \\
Mode(s)     &  (MeV)             &  (MeV)       & (GeV)         & $N$ \\
\hline
$\pi^+\pi^-/K^+\pi^-/K^+K^-$ & $\pm 25$    & $-185$ $<\Delta E< 140$
& $5.21$ $< M_B < 5.30$ &  453 \\
$\pi^+\pi^0/K^+\pi^0$        & $+43/{-55}$ & $\pm 300$
& $5.20$ $< M_B < 5.30$   &  896 \\
$\pi^0\pi^0$                 & $+51/{-85}$ & $\pm 300$
& $5.20$ $< M_B < 5.30$  &  104 \\
$K^0\pi^0$                   & $+44/{-53}$ & $\pm 200$
& $5.20$ $< M_B < 5.30$  &   44 \\
$K^0\pi^+$                   & $\pm 27$    & $\pm 200$
& $5.20$ $< M_B < 5.30$  &  220 \\
\end{tabular}
\label{tab:mlfits}
\end{center}
\end{table}

\begin{table}
\mediumtext
\begin{center}
\caption{Results from the likelihood analyses: the signal yield central
value from the fit ($N_S$), detection efficiencies (${\cal E}$),
the 90\% confidence level
upper limit on the number of signal events ($N_{\rm UL}$), the 90\% CL
upper limit of the
branching fraction (UL $\cal B$), and the theoretical predictions
for the branching fractions  \protect\cite{deandrea,chau,trampetic}.
We also include the measured branching fraction ($\cal B$) for the sum of
$\pi^+\pi^-$ and $K^+\pi^-$, where the first error is statistical,
the second is the systematic error from the yield, and the third is
the systematic error from the efficiency. }
\begin {tabular}{l c r c d d c}
    &  &\multicolumn{1}{c}{${\cal E}$} &$\cal B$   &
& UL $\cal B$ &Theory      \\
Mode&   $N_S$ &\multicolumn{1}{c}{(\%)} &($10^{-5}$)
&  $N_{\rm UL}$& ($10^{-5}$) &  ($10^{-5}$)\\
\hline
$h^+\pi^-$   &17.2$^{+5.6+2.2}_{-4.9-2.5}$     &$37\pm3$
&  $1.8^{+0.6+0.2}_{-0.5-0.3}\pm0.2$ &         &    & \\
{}~~~~$\pi^+\pi^-$ &9.4$^{+4.9}_{-4.1}$          &$37\pm3$
&               & 17.9    &2.0  & 1.0--2.6	\\
{}~~~~$K^+\pi^-$   &7.9$^{+4.5}_{-3.6}$          &$37\pm3$
&               & 15.3    &1.7  & 1.0--2.0        \\
{}~~~~$K^+K^-$     &0.0$^{+0.8}_{-0.0}$          &$37\pm3$
&               &  3.5    &0.40 &	--              \\
$K^+\pi^0$   &4.9$^{+3.6}_{-2.8}$          &$33\pm3$
&               & 11.2    &1.4  & 0.3--1.3         \\
$\pi^+\pi^0$ & 5.0$^{+4.2}_{-3.2}$         &$33\pm3$
&               & 13.1    &1.7  & 0.6--2.1        \\
$\pi^0\pi^0$ & 1.2$^{+1.7}_{-0.9}$         &$26\pm4$
&               &5.2      &0.91 & 0.03--0.10    \\
$K^0\pi^+$   & 5.2$^{+3.5}_{-2.8}$         &$11\pm2$
&               &11.3     &4.8  & 1.1--1.2               \\
$K^0\pi^0$   & 2.3$^{+2.2}_{-1.5}$         &$7\pm1$
&               &6.2      &4.0  & 0.5--0.8         \\
\end {tabular}
\label{tab:mlresults}
\end{center}
\end {table}

\begin{table}
\mediumtext
\begin{center}
\caption{Dominant variations in the upper limit signal yield (\%)
due to systematic uncertainties in the fit shapes.}
\vspace{0.5cm}
\begin{tabular}{l c c c c c c}
Mode & Background $M_B$\ & Signal $M_B$\ & Signal $\Delta E$\
& $\cal F$ & $dE/dx$\ & Total \\
\hline
$\pi^+\pi^-$ & $8.6$ & $2.9$ & $5.4$ & $4.7$ & $5.1$ & $13$\\
$K^+\pi^-$ & $5.0$ & $3.3$ & $2.6$ & $2.5$ & $4.6$ & $8$\\
$K^+K^-$ & $4.7$ & $<1$ & $<1$ & $2.2$ & $3.2$ & $9$\\
$K^+\pi^0$ & $5.9$\ & $<0.5$\ & $2.3$\ & $2.0$ & $2.7$ & $7$\\
$\pi^+\pi^0$ & $13.5$\ & $2.8$\ & $3.3$\ & $3.2$ & $1.7$ & $15$\\
$\pi^0\pi^0$ & $12.1$ & $3.1$ & $1.2$ & $5.1$ & $-$ & $14$\\
$K^0\pi^+$   & $7.5$ & $4.8$ & $1.8$ & $5.7$ & $-$ & $9$\\
$K^0\pi^0$ & $6.9$ & $1.0$ & $1.1$ & $1.6$ & $-$ & $7$\\
 & & & & & & \\
\tablenotemark[1] $N_{\rm sum}$
& $+10.4/{-12.8}$\ & $+3.9/{-3.5}$\   & $+1.5/{-0.7}$
& $+5.9/{-6.4}$\ & $\pm1$\ & $+13/{-15}$ \\
\tablenotemark[1] $R$ & $+2.0/{-3.3}$ & $+1.7/{-1.4}$
& $\pm 7.2$ & $+1.7/{-2.6}$ & $\pm 5.6$ & $+9/{-10}$\\
\end{tabular}
\tablenotemark[1]{Systematic errors on central value.}
\label{tab:mlsyst}
\end{center}
\end{table}
\clearpage
\begin {table}
\narrowtext
\begin{center}
\caption{Results of the event-counting analyses: the number of
events in the signal region ($N_S$), the estimated background in the
signal region ($N_B$), the
90\% confidence level upper limit on the branching fractions
(UL $\cal B$), and theoretical
predictions \protect\cite{deandrea,chau,trampetic}.  Indicated in parenthesis
are the $K^*$ decay modes used.}
\vspace{0.3cm}
\begin {tabular}{l c c c c }
	&        &       & UL $\cal B$    & Theory \\
{Mode}  & $N_S$  & $N_B$ & $(10^{-5})$    & ($10^{-5}$) \\
\hline
$\pi^{\pm}\rho^{\mp}$  & 7       &$2.9\pm0.7$ &8.8 & 1.9--8.8\\
$\pi^0\rho^0$ 	       & 1       &$1.8\pm0.6$ &2.4 & 0.07--0.23\\
$\pi^+\rho^0$          & 4       &$2.3\pm0.3$ &4.3 & 0.0--1.4\\
$\pi^0\rho^+$          & 8       &$5.5\pm1.2$ &7.7 & 1.5--3.9\\
                       &         &            &    &  \\
$K^+\rho^-$ & 2       &$2.0\pm0.4$ &3.5 & 0.0--0.2 \\
$K^0\rho^0$ & 0       &  0         &3.9 & 0.004--0.04 \\
$K^+\rho^0$ & 1       &$3.8\pm0.2$ &1.9 & 0.01--0.06\\
$K^0\rho^+$ & 0       &  0         &4.8 & 0--0.03         \\
                       &         &            &    &  \\

$K^{*+}\pi^-$           & 3       &$0.7\pm0.2$ & 7.2& 0.1--1.9 \\
{}~~~~$ (K^+\pi^0)$     & ~~~~~~(3)       &~~~~~~$(0.7\pm0.2)$ &    &  \\
{}~~~~$(K^0\pi^+)$	& ~~~~~~(0)       &~~~~~~(0)         &    &  \\

$K^{*0}\pi^0$          &    0    & $1.1\pm0.3$   &2.8 & 0.3--0.5 \\
{}~~~~$ (K^+\pi^-)$    & ~~~~~~(0)    &~~~~~~$(1.1\pm0.3)$ &    &      \\

$K^{*+}\pi^0$          & 4    & $1.9\pm0.7$ &9.9  &0.05--0.9        \\
{}~~~~$ (K^+\pi^0)$    & ~~~~~~(3)    & ~~~~~~$(1.9\pm0.7)$ &     &   \\
{}~~~~$ (K^0\pi^+)$    & ~~~~~~(1)    & ~~~~~~(0)            &     &  \\

$K^{*0}\pi^+$          & 2    &$1.0\pm0.6$ &4.1 & 0.6--0.9\\
{}~~~~$ (K^+\pi^-)$    & ~~~~~~(2)    &~~~~~~$(1.0\pm0.6)$ &    &     \\
{}~~~~$ (K^0\pi^0)$    & ~~~~~~(0)    &~~~~~~(0)            &    &    \\

                       &         &            &    &  \\
$K^0\phi$     & 1    &    0       &8.8 & 0.07--1.3\\
$K^+\phi$     & 0    &    0       &1.2 & 0.07--1.5  \\

$K^{*0}\phi$         & 2     &0            &4.3 &  0.02--3.1   \\
{}~~~~($K^+\pi^-$) &~~~~~~ (2)     &~~~~~~(0)            &    &     \\
{}~~~~($K^0\pi^0$)  &~~~~~~ (0)     &~~~~~~(0)            &     &	\\

$K^{*+}\phi$       & 1     &0            & 7.0  & 0.02--3.1     \\
{}~~~~($K^+\pi^0$) &~~~~~~ (0)     &~~~~~~(0)            &   &     \\
{}~~~~($K^0\pi^+$) &~~~~~~ (1)     &~~~~~~(0)	    &   &     \\

$\phi\phi$         &  0    & 0          & 3.9 &            	\\

\end {tabular}
\label{tab:evcount_results}
\end{center}
\end {table}

\begin{table}
\begin{center}
\caption{Reconstruction efficiencies (${\cal E}_r$),
the products of the appropriate daughter branching fractions (${\cal E}_d$),
and total detection efficiencies
(${\cal E}\equiv {\cal E}_r\times {\cal E}_d$) for the
event-counting analyses. }
\vspace{0.5cm}
\begin{tabular}{l c d c}
Mode & ${\cal E}_r$ (\%) & ${\cal E}_d$ & ${\cal E}$ (\%)\\
\hline
$\pi^\pm\rho^\mp$  & $5.3 \pm 1.1$ & 0.988   & $5.2\pm1.0$ \\
$\pi^0\rho^0$      & $6.5 \pm 1.3$ & 0.988   & $6.4\pm1.2$ \\
$\pi^+\rho^0$      & $7.4 \pm 1.5$ &  1.0    & $7.4\pm1.5$ \\
$\pi^0\rho^+$      & $5.5 \pm 1.1$ & 0.976   & $5.4\pm1.1$ \\
 & & & \\
$K^+\rho^-  $      & $5.7 \pm 1.1$ & 0.988   & $5.7\pm1.1$ \\
$K^0\rho^0  $      & $7.8 \pm 1.2$ & 0.343   & $2.7\pm0.4$ \\
$K^+\rho^0  $      & $7.1 \pm 1.4$ &  1.0    & $7.1\pm1.4$ \\
$K^0\rho^+  $      & $6.4 \pm 1.0$ & 0.339   & $2.2\pm0.3$ \\
 & & & \\

$K^{*+}\pi^-$      &               &         & $3.7\pm0.4$ \\
{}~~~~$(K^+\pi^0)$   & $4.5 \pm 0.8$ & 0.329   & ~~~~~~$(1.5\pm0.2)$ \\
{}~~~~$(K^0\pi^+)$   & $9.8 \pm 2.0$ & 0.228   & ~~~~~~$(2.2\pm0.3)$ \\
$K^{*0}\pi^0$      &               &         & \\
{}~~~~$(K^+\pi^-)$   &$6.1 \pm 1.2$  & 0.657   & $4.0\pm0.8$ \\
$K^{*+}\pi^0$      &               &         & $3.0\pm0.4$ \\
{}~~~~$(K^+\pi^0)$   & $3.9 \pm 0.8$ & 0.325   & ~~~~~~$(1.3\pm0.2)$ \\
{}~~~~$(K^0\pi^+)$   & $7.6 \pm 1.5$ & 0.226   & ~~~~~~$(1.7\pm0.3)$ \\
$K^{*0}\pi^+$      &               &         & $5.6\pm0.9$  \\
{}~~~~$(K^+\pi^-)$   & $7.1 \pm 1.4$ & 0.665   & ~~~~~~$(4.7\pm0.9)$ \\
{}~~~~$(K^0\pi^0)$   & $7.9 \pm 1.6$ & 0.113   & ~~~~~~$(0.9\pm0.2)$ \\
 & & & \\

$K^0\phi$	   &$11.9\pm1.8$   & 0.168   & $2.0\pm0.3$ \\
$K^+\phi$	   &$17.8\pm2.7$   & 0.491   & $8.7\pm1.3$ \\
$K^{*0}\phi$       &               &         & $5.6\pm0.8$ \\
{}~~~~~~$(K^+\pi^-)$ &$16.2\pm2.4$   &0.327    & ~~~~~~$(5.3\pm0.8)$ \\
{}~~~~~~$(K^0\pi^0)$ &$ 5.6\pm0.8$   &0.055    & ~~~~~~$(0.3\pm0.1)$ \\
$K^{*+}\phi$       &               &         & $2.6\pm0.5$ \\
{}~~~~~~$(K^+\pi^0)$ &$ 9.2\pm1.4$   &0.162    & ~~~~~~$(1.5\pm0.2)$ \\
{}~~~~~~$(K^0\pi^+)$ &$ 9.4\pm1.4$   &0.112    & ~~~~~~$(1.1\pm0.2)$ \\
$\phi\phi$	   &$11.0\pm1.7$   &0.241    & $2.7\pm0.4$ \\
\end{tabular}
\label{tab:eff}
\end{center}
\end{table}

\clearpage

%\end{document}

%\clearpage

%\clearpage

%\clearpage

%\clearpage

%\clearpage

%\clearpage


\begin{references}
\bibitem{chargeconjugation}
Throughout this paper, charge conjugate states are implied.
\bibitem{kstars}
In the mode $B^0 \rightarrow K^{*0}\pi^0$, we only use the $K^{*0}$
decay mode $K^{*0} \rightarrow K^+\pi^-$.
\bibitem{thrustnotflat}
Because of non-ideal matching of shower fragments to the
candidate tracks that produced them, the $\cos \theta_T$ distribution
for signal
shows very slight peaking at $\pm 1$ for modes with high momentum
charged tracks.
\bibitem{chargestrangeness}
Charge-strangeness
correlations do not allow the unlike-sign track to be a kaon.
\end{references}
\end{document}